\documentclass{jnmp01b}

\usepackage{amsmath}

\numberwithin{equation}{section}

\begin{document}

\renewcommand{\evenhead}{I.A.B. Strachan}
\renewcommand{\oddhead}{Jordan Manifolds and Dispersionless KdV Equations}

% Title

\thispagestyle{empty}

\begin{flushleft}
\footnotesize \sf
Journal of Nonlinear Mathematical Physics \qquad 2000, V.7, N~4,
\pageref{firstpage}--\pageref{lastpage}.
\hfill {\sc Article}
\end{flushleft}

\vspace{-5mm}

\copyrightnote{2000}{I.A.B. Strachan}

\Name{Jordan Manifolds and Dispersionless KdV Equations}

\label{firstpage}

\Author{I.A.B. STRACHAN}

\Adress{Department of Mathematics, University of Hull, Hull, HU6 7RX, England\\
E-mail: i.a.b.strachan@hull.ac.uk}

\Date{Received March 13, 2000; Revised June 9, 2000;
Accepted June 26, 2000}

\begin{abstract}
\noindent
Multicomponent KdV-systems are defined in terms of a set of
structure constants and, as shown by Svinolupov, if these define a Jordan algebra the
corresponding equations may be said to be integrable, at least in the sense of having
higher-order
symmetries, recursion operators and hierarchies of conservation laws. In this paper
the dispersionless limits of these
Jordan KdV equations are studied, under the assumptions that the Jordan algebra has a unity
element and a compatible non-degenerate inner product. Much of this structure may be
encoded in a so-called Jordan manifold, akin to a Frobenius manifold. In particular
the Hamiltonian
properties of these systems are investigated.
\end{abstract}

% The article

\section{Introduction}

In \cite{[Sv]} Svinolupov studied $N$-component KdV-type equations of the form
\begin{equation}
u^i_t = u^i_{xxx} + a_{jk}^{~~i} u^j u^k_x\,,\quad i\,,j\,,k=1\,\ldots\,,N\,,
\label{jordankdv}
\end{equation}
where the fields $u^i$ depend on $x$ and $t$ alone and the $a_{jk}^{~~k}$ are
constants, symmetric in the lower indices. In particular, necessary and sufficient
conditions were found for (\ref{jordankdv}) to possess higher symmetries and conservation laws,
these being best expressed in terms of the algebra defined by the constants $a_{jk}^{~~i}\,.$

Let ${\cal F}$ be a finite dimensional commutative algebra over $\mathbb{C}$ with basis
$e_i\,,\,i=1\,,\ldots\,,N\,,$ with multiplication
\[
e_i \circ e_j = a_{ij}^{~~k} e_k\,.
\]
The algebra is said to be a Jordan algebra if $(x^2\circ y)\circ x = x^2\circ(y\circ x)$
for all $x\,,y\in{\cal F}\,.$ This is a cubic conditions on the structure constants
$a_{ij}^{~~k}\,.$ Note that if the algebra is associative then it is automatically a Jordan
algebra. In what follows it will be useful to introduce the so-called
associator $\Delta_{ijk}^{~~~s}$ defined by
\[
(e_i \circ e_j) \circ e_k - e_i \circ (e_j \circ e_k)
=\Delta_{ijk}^{~~~s}e_s\,,
\]
or, in components, by
\[
\Delta_{ijk}^{~~~s} = a_{ij}^{~~r} a_{rk}^{~~s} - a_{jk}^{~~r}
a_{ir}^{~~s}\,.
\]
This is a measure of the deviation of the Jordan algebra from being associative. With this the
Jordan condition may be written succinctly as
\[
a_{(ij}^{~~~r} \Delta_{k)mr}^{~~~~n} = 0\,.
\]
More details of Jordan algebras may be found in \cite{[Sv]} and in \cite{[Sc]}. In this paper two additional conditions will be assumed. For simple, irreducible, Jordan
algebras these hold automatically.

\begin{itemize}

\item[$\bullet$] The algebra has a unity element $e_1$ such that
\begin{equation}
e_1 \circ e_i = e_i \,, \quad\quad\forall i\,.
\label{unity}
\end{equation}
This condition is very weak, since any algebra without a unity element may be appended with
a unity element \cite{[Sc]};

\item[$\bullet$] Let $M(x)$ denote the operation of multiplication by the element $x$ in
${\cal F}\,, M(x) y = x \circ y\,.$ With this one may define a canonical bilinear symmetric form
\[
<x,y> = {\rm tr} \Big\{ M( x \circ y )\Big\}  \,,\quad\quad \forall x\,,y \in {\cal F}\,.
\]
In local coordinates this will be denoted $\eta_{ij}\,,$ so
\begin{equation}
\eta_{ij} = a_{ij}^{~~k} a_{km}^{~~m}\,.
\label{defeta}
\end{equation}
The assumption that will be made is that this bilinear form is non-degenerate and that
is satisfies the Frobenius condition
\begin{equation}
< a \circ b , c> = < a , b \circ c>\,, \quad \forall a\,,b\,,c \in {\cal F}\,.
\label{conditionA}
\end{equation}

\end{itemize}
With these conditions the systems (\ref{jordankdv}) possesses an infinite series of non-trivial canonical
conservation laws as well as higher symmetries and a recursion operator \cite{[Sv]}. Hence it may be
regarded as an integrable system.

In purpose of this paper is to study the dispersionless limit of these systems, namely
\begin{equation}
u^i_t = a_{jk}^{~~i} u^j u^k_x\,,\quad i\,,j\,,k=1\,\ldots\,,N\,.
\label{djordankdv}
\end{equation}
These systems are examples of equations of hydrodynamic type, that is, they are of the form
\begin{equation}
u^i_t = v^i_j({\bf u})\,u^j_x\,.
\label{genhydro}
\end{equation}
In particular the conservation laws and bi-Hamiltonian structures
for these systems will be studied. The system (\ref{djordankdv}) may be rewritten in terms
of an ${\cal F}$-valued function
\[
{\cal U}(x,t) = u^i(x,t)\, e_i
\]
as
\begin{equation}
{\cal U}_t = {\cal U} \circ {\cal U}_x\,.
\label{fvalued}
\end{equation}
The main, and historically the first, examples of Jordan algebras came from
matrix multiplication, here denoted $\cdot$ via
\[
a \circ b = \frac{1}{2} (a \cdot b + b \cdot a)\,.
\]
Such Jordan algebras are said to be special, though it is important to note that not all
Jordan algebras arise in this way.
For such systems (\ref{fvalued}) simplifies further to
\begin{equation}
{\cal U}_t = \frac{1}{2} \big( {\cal U}^2 \big)_x\,.
\label{matrixhopf}
\end{equation}
This equation has appeared in the literature before \cite{[F1]} and may
be thought of as a matrix Hopf equation. The simplest example,
where ${\cal U}$ is a $2\times 2$ symmetric matrix corresponds to
the $D_3$-Jordan algebra constructed below. Such a matrix
equations are integrable, in the sense that they arise as the
compatibility conditions for the Lax pair
\begin{gather*}
\left[\partial_x - \lambda \,{\cal U}{\phantom{^2}}  \right] \phi  =  0 \,,
\\[1ex]
\left[\partial_t - \lambda \,{\cal U}^2\right] \phi  =  0
\end{gather*}
and so can be integrated by the inverse scattering transform.

%The Jordan condition will not play a major role in this paper; many of the results
%will hold for any commutative algebra with a non-degenerate inner product satisfying
%the conditions (\ref{unity}) and (\ref{conditionA}). The main reason for the assumption
%is to ensure that the systems (\ref{djordankdv}) are the dispersionless limits of the
%integrable, dispersive, systems (\ref{jordankdv}), where by Svinolupov's result one
%requires the Jordan condition. As is well known, there are many more dispersionless
%integrable systems than occur as dispersionless limits of dispersive integrable systems.
%Presumably the Jordan condition would come into play as a condition for the consistent
%deformation of the dispersionless systems \cite{[EYY]}.

Under certain natural assumptions it may be shown that equation
(\ref{djordankdv}) has an infinite hierarchy of hydrodynamic
conservation laws if and only if the corresponding algebra is
Jordan \cite{[St]}. The resultant systems are thus the
dispersionless limits of the integrable, dispersive, systems
(\ref{jordankdv}) introduced by Svinolupov \cite{[Sv]}.

\section{Jordan manifolds}
\resetfootnoterule

Condition (\ref{conditionA}) may be written in local coordinates
as
\[
\eta_{in} a_{jk}^{~~n} = \eta_{kn} a_{ji}^{~~n}\,.
\]
Using $\eta_{ij}$ and $\eta^{ij} = (\eta_{ij})^{-1}$ to lower and raise indices, this condition,
together with the commutativity of the Jordan multiplication, implies that
\[
a_{ijk} = \eta_{kn} a_{ij}^{~~n}
\]
is totally symmetric. This in turn implies the existence of a scalar function $F\,,$ which will be called
the prepotential, such that
\[
a_{ijk}=\frac{ \partial^3 F}{\partial u^i \partial u^j \partial u^k}\,,
\]
there being no integrability conditions since $a_{ijk}$ are constants. This prepotential
is defined only up to quadratic terms, and satisfies the normalization condition
\[
\eta_{ij}=a_{1ij}\,.
\]
This follows from the existence of the unity element in the algebra
(so $a_{1i}^{~~j}=\delta_i^j$) and definition (\ref{defeta}).

These conditions bear a striking resemblance to the definition of a Frobenius
manifold \cite{[D]}, and in the case where the Jordan multiplication is actually
associative, they are identical. Thus one is lead to the definition of a Jordan manifold.

\bigskip

\noindent{\bf Definition}. Let $F=F({\bf u})\,, {\bf u}=(u^1\,,\ldots\,, u^N)$ be a scalar
function such that the third derivatives
\[
a_{ijk} = \frac{ \partial^3 F}{\partial u^i \partial u^j \partial u^k}
\]
satisfy the following conditions:

\begin{itemize}

\item{[Normalization]} $\eta_{ij} = a_{1ij}$ is a constant non-degenerate matrix. This, together
with its inverse, may be used to lower and raise indices, and hence may be regarded as a (flat)
metric on the manifold;

\item{[Jordan condition]} The structure constants
\[
c_{ij}^{~~k} = c_{ijr} \eta^{rk}
\]
define a Jordan algebra
\begin{equation}
\frac{\partial~}{\partial u^i} \circ \frac{\partial~}{\partial u^j}
= a_{ij}^{~~k} \frac{\partial~}{\partial u^k}
\label{tangentmult}
\end{equation}
on each tangent space of the manifold. The unity vector field $\frac{\partial~}{\partial u^1}$
will be denoted by $e\,.$

\end{itemize}

\noindent It is not known whether there are non-trivial, i.e. non-cubic, examples of the
function $F\,,$ the Jordan condition, a system of third order, third degree, partial differential
equations, being far more rigid than the WDVV-equations of associativity.

For the purposes of this paper the $c_{ijk}$ must be constant, so $F$ must be a cubic function
(ignoring quadratic terms which do not effect these constants). Thus $F$ must be a
homogeneous function, and hence one may add an extra condition to the above definition:

\begin{itemize}

\item{[Homogeneity]} The prepotential $F$ must be a homogeneous function,
\[
{\cal L}_E F = 3 F
\]
where ${\cal L}_E$ is the Lie-derivative along the Euler vector field
\[
E=u^i \frac{ \partial~~}{\partial u^i}\,.
\]

\end{itemize}

\noindent There is considerable scope for investigating Jordan manifolds in more generality
than will be needed here, where only cubic $F$'s will be considered.

\bigskip

\noindent{\bf Example}. In two dimensions there are, up to isomorphism, five Jordan algebras
\cite{[Sc],[Sv]}.
The only one to have a unity element is also associative, corresponding to the case $N=2$ in the
example below.

\bigskip

\noindent{\bf Example}. Let
\[
F=\frac{1}{2} (u^1)^2 u^N + \frac{1}{2} u^1 \sum_{i+j=N+1\,: i\,,j>1} u^i u^j\,.
\]
The resulting algebra is associative, describing classical, rather than quantum, cohomology.
The resulting dispersionless KdV system is:
\[
\left(\begin{array}{c}
u^1 \\ u^2 \\ u^3 \\ \vdots \\ u^{N-1} \\ u^N
\end{array}\right)_t = 2
\left(\begin{array}{cccccc}
u^1 &     &     &       &        &      \\
u^2 &  u^1&    &     &   {\bf 0}      &      \\
u^3 &  u^2& u^1 &        &        &    \\
\vdots & \vdots & \vdots & \ddots &  & \\
u^{N-1} & u^{N-2} & u^{N-3} & \ldots & u^1 & \\
u^N     & u^{N-1} & u^{N-2} & \ldots & u^2 & u^1
\end{array}\right)\,
\left(\begin{array}{c}
u^1 \\ u^2 \\ u^3 \\ \vdots \\ u^{N-1} \\ u^N
\end{array}\right)_x\,.
\]
where all the upper diagonal entries are zero.

\bigskip

\noindent{\bf Example}. (The $D_N$-Jordan algebras \cite{[Sc],[Sv]}) Let the Jordan product on $\cal F$ be
defined by
\[
x \circ y = (a,x) y + (a,y) x - (x,y) a\,,
\]
where $dim{\cal F}>2$ and $(~,~)$ is the ordinary inner product with $(a,a)\neq 0\,.$ Without
loss of generality, one may take $a=e_1$ and hence obtain the multiplication table
\begin{gather*}
e_1 \circ e_i  =  + e_i \,, \\
e_i \circ e_i  =  - e_1 \,, \quad i=2\,,\ldots\,, N\,, \\
e_i \circ e_j  =  0 \,,\quad {\rm otherwise}\,.
\end{gather*}
(Such a multiplication comes from an underlying Clifford multiplication). The structure
constants may be written compactly as
\begin{equation}
a_{ij}^{~~k} = \delta_{1i} \delta_{jk} + \delta_{1j} \delta_{ik} - \delta_{1k} \delta_{ij}\,.
\label{dnconstants}
\end{equation}
The inner product is given by (on ignoring an overall factor of $N$)
\[
\eta_{ij} = {\rm diag}\,(+1\,,-1\,,-1\,,\ldots\,, -1)
\]
and the prepotential $F$ is
\[
F = \frac{1}{6} (u^1)^3 - \frac{1}{2} u^1 \sum_{i=2}^N (u^i)^2\,.
\]
The corresponding dispersionless KdV system may easily be written down; for $N=3$ this
being:
\begin{equation}
\begin{split}
&u_t  =  -3 (u^2-v^2-w^2)_x\,, \\
&v_t  =  -6 (uv)_x\,,\\
&w_t  =  -6 (uw)_x\,.
\end{split}
\label{basicexample}
\end{equation}
This is the only irreducible Jordan algebra in dimension 3.

\bigskip

\noindent{\bf Example}. A canonical class of Jordan algebras are defined by
the multiplication
\[
x \circ y = \frac{x\cdot y + y\cdot x}{2}\,,
\]
where $x$ and $y$ are matrices and $\cdot$ is matrix multiplication. Consider the
(irreducible) algebra of symmetric $3\times 3$ matrices with elements in $\mathbb{C}\,.$ Taking
as a basis
\[
\begin{array}{@{}lll}
e_1 = \left(\begin{array}{rrr} +1 & 0 & 0 \\ 0 & +1 & 0 \\ 0 & 0 & +1 \end{array}\right) &
e_2 = \left(\begin{array}{rrr} +1 & 0 & 0 \\ 0 & -1 & 0 \\ 0 & 0 & +1 \end{array}\right) &
e_3 = \left(\begin{array}{rrr} +1 & 0 & 0 \\ 0 & +1 & 0 \\ 0 & 0 & -1 \end{array}\right) \\
&&\\
e_4 = \left(\begin{array}{rrr} 0 & +1 & 0 \\ +1 & 0 & 0 \\ 0 & 0 & 0 \end{array}\right) &
e_5 = \left(\begin{array}{rrr} 0 & 0 & +1 \\ 0 & 0 & 0 \\ +1 & 0 & 0 \end{array}\right) &
e_6 = \left(\begin{array}{rrr} 0 & 0 & 0 \\ 0 & 0 & +1 \\ 0 & +1 & 0 \end{array}\right)
\end{array}
\]
one may drive the structure constants of the Jordan algebra and hence the inner product.
The resulting prepotential is
\[
F=\frac{1}{360}
\left[
\begin{array}{c}
+3\,{{u_1}^3} + {{u_2}^3} + 3\,{{u_2}^2}\,u_3 + {{u_3}^3} +
     3\,{{u_1}^2}\,\left( u_2+ u_3 \right) + 6\,u_3\,{{u_4}^2}\\
   +  3\,u_2\,\left( {{u_3}^2} + 2\,{{u_5}^2} \right)
   + 6\,u_4\,u_5\,u_6\\
   + 3\,u_1\,      \left( 3\,{{u_2}^2} - 2\,u_2\,u_3 +
3\,{{u_3}^2} + 2\,{{u_4}^2} + 2\,{{u_5}^2} + 2\,{{u_6}^2} \right)
\end{array}
\right]\,,
\]
where $u_i=u^i$ for notational convenience. The metric is not diagonal in the basis. It may, of
course, be diagonalized, but this will change the simple form of the above matrix-basis.

\section{Conservation laws and Hamiltonian structures}

Equation (\ref{djordankdv}) may be written in Hamiltonian form
\[
u^i_t = \eta^{ij} \frac{d~}{dx} \frac{ \delta~}{\delta u^j}
\left(\frac{a_{pqr}u^p u^q u^r}{3!}\right)\,.
\]
The Hamiltonian densities form an infinite series of conservation laws, which generate
an infinite family of commuting flows. These densities will be labelled by their degree,
the first two being
\begin{gather*}
h^{(2)}  =  \frac{1}{2!} \eta_{pq} u^p u^q\,, \\
h^{(3)}  =  \frac{1}{3!} a_{pqr}u^p u^q u^r\,,
\end{gather*}
These define commuting flows (defined so $t_1$ may be identified with $x$)
\[
u^i_{t_{n}} = {\cal H}^{ij}_{(1)} \frac{\delta h^{(n+1)}}{\delta u^j}\,,
\]
where ${\cal H}_{(1)}$ is the operator
\[
{\cal H}^{ij}_{(1)} = \eta^{ij} \frac{d~}{dx}\,.
\]
The existence of a unity element in the algebra implies that these densities
are connected by the relation
\[
\frac{\partial h^{(n+1)}}{\partial u^1} = {\rm constant~} h^{(n)}\,.
\]
\begin{proof}
Any hydrodynamic conservation law
\[
Q[{\bf u}]_t = {\rm Flux}[ {\bf u} ]_x
\]
may be expanded, using (\ref{djordankdv}), yielding
\[
\frac{\partial {\rm \,Flux}}{\partial u^k} = a_{jk}^{~~i} u^j \frac{\partial Q}{\partial u^i}\,.
\]
The integrability condition for this is
\begin{equation}
a_{jk}^{~~i} u^j \frac{ \partial^2 Q}{\partial u^i \partial u^p} =
a_{jp}^{~~i} u^j \frac{ \partial^2 Q}{\partial u^i \partial u^k}.
\label{conservationlaw}
\end{equation}
By differentiating this with respect to $u^1\,,$ the unity element,
one finds that $\partial Q/\partial u^1$ satisfies
the same equation, and hence is also conserved. These conserved densities
are all homogeneous and may be labelled by their degree, so by Euler's theorem
\begin{equation}
u^i \frac{\partial h^{(n)}}{\partial u^i} = n h^{(n)}.
\label{euler}
\end{equation}
They may also be normalised so that
\begin{equation}
\frac{\partial h^{(n)}}{\partial u^1} = h^{(n-1)}\,.
\label{recursiondown}
\end{equation}
The basic relation (\ref{conservationlaw}) may also be used to derive a
recursion relation amongst the densities. Let $p=1$ in (\ref{conservationlaw}),
so, on using the unity relation, (\ref{euler}) and (\ref{recursiondown})
\begin{equation}
a_{jk}^{~~i} u^j \frac{\partial h^{(n-1)}}{\partial u^i}   =
(n-1) \frac{\partial h^{(n)}}{\partial u^k}\,.
\label{usefulequation}
\end{equation}
Multiplying by $u^k$ and using Euler's theorem (\ref{euler}) again yields
\begin{equation}
h^{(n)} = \frac{1}{n(n-1)} a_{jk}^{~~i} u^j u^k \frac{\partial h^{(n-1)}}{\partial u^i}\,.
\label{recursion}
\end{equation}
Alternatively, from (\ref{usefulequation}) and Euler's theorem one may derive
\[
\Bigg[
\frac{\partial^2 h^{(n)}}{\partial u^i\partial u^j} -
c_{ij}^{~~k} \frac{\partial h^{(n-1)}}{\partial u^k}\Bigg] u^j =0\,.
\]
The term in square brackets will not, in general, be zero, but for Frobenius
manifolds it does vanish, while in the examples below it does not.
Thus one has the following recursion scheme for these Hamiltonian
densities:
\begin{equation*}
\eta_{1i} u^i = h^{(1)}\rightleftarrows
\cdots \rightleftarrows h^{(n)} \begin{array}{c}
{}_{(\ref{recursion})}
\\ \rightleftarrows \\ {}^{(\ref{recursiondown})} \end{array} h^{(n+1)}
\rightleftarrows \cdots\,. \tag*{\qed}
\end{equation*}
\renewcommand{\qed}{}
\end{proof}

\bigskip

\noindent{\bf Example}. The matrix Hopf equation

\medskip

The matrix Hopf equation (\ref{matrixhopf}) is known to have
the conservation laws
\[
h^{(n)}=\frac{1}{n!} Tr\left({\cal U}^n\right)\,,
\]
where $Tr({\cal V})=<{\cal V},e_1>\,,$ since $e_1$ is just the
identity matrix. These coincide with the conservation laws
previously constructed, for example

\begin{align*}
h^{(3)} & =  \frac{1}{6} c^k_{ij} c^s_{kr} u^i u^j u^r
<e_s,e_1>\,,\\
& =  \frac{1}{6} c^k_{ij} \eta_{kr} u^i u^j u^r \,, \\
& =  \frac{1}{6} c_{ijk} u^i u^j u^k \,.
\end{align*}
To show that that these traces (\ref{recursiondown}) is
straightforward:
\begin{align*}
\frac{\partial h^{(n+1)}}{\partial u^1} & =  \frac{1}{(n+1)!}
\frac{\partial~}{\partial u^1} Tr\left({\cal U}^{n+1}\right) \,,
\\
& =  \frac{1}{n!} Tr \left( {\cal U}^n . e_1\right)
 =\frac{1}{n!} Tr \left( {\cal U}^n \right)\,,\\
& =  h^{(n)}\,,
\end{align*}
and to show that they satisfy (\ref{usefulequation})(and
hence (\ref{recursion}) by homogeneity) is
similar:
\[
\frac{\partial h^{(n)}}{\partial u^k} = \frac{1}{(n-1)!}
Tr\left[ {\cal U}^{n-1} ~e_k\right]
\]
so
\begin{align*}
a^i_{jk} u^j\frac{\partial h^{(n)}}{\partial u^i}
&=
a^i_{jk} u^j \frac{1}{(n-1)!} Tr\left[
{\cal U}^{n-1} e_i\right]\\
& =
\frac{1}{(n-1)!} Tr\left[ a^i_{jk} u^j e_i {\cal U}^{n-1}
\right]\\
& =  \frac{1}{(n-1)!} Tr\left[ {\cal U}
\frac{\partial{\cal U}}{\partial u^k} {\cal U}^{n-1}\right]\\
& =  n \frac{\partial h^{(n+1)}}{\partial u^k}\,.
\end{align*}
Hence these trace formulae for conserved quantities satisfy
the general results (\ref{recursiondown}) and
(\ref{recursion})\,.

\bigskip

\noindent{\bf Example}. (The $D_3$-Jordan algebra)

The first few conserved densities are:
\begin{gather*}
h^{(2)}  =  \frac{1}{2!} \Big\{ u^2-v^2-w^2 \Big\} \,, \\
h^{(3)}  =  \frac{1}{3!} \Big\{ u^3 - 3u(v^2+w^2) \Big\} \,, \\
h^{(4)}  =  \frac{1}{4!} \Big\{ u^4 - 6 u^2 (v^2+w^2) + (v^2+w^2)^2 \Big\}\,.
\end{gather*}
The general terms may easily by derived:
\[
h^{(n)} = \frac{1}{n!} \sum_{r=0}^{ \big[ \frac{n}{2} \big] }
(-1)^r \left(\begin{array}{c} n \\ 2r \end{array}\right) (v^2+w^2)^r \,u^{(n-2r)}\,.
\]
These may be amalgamated into a generating function, the coefficients of $\lambda$ in the power
series expansion being the conserved densities,
\[
{\cal Q}(\lambda) = e^{\lambda u} \cos \lambda \sqrt{v^2+w^2}\,.
\]
Similiarly there is a second family of conservation laws given by the generating function
\[
{\cal Q}(\lambda) = e^{\lambda u} \sin \lambda \sqrt{v^2+w^2}\,.
\]
These may be combined as
\[
{\cal Q}(\lambda) = e^{\lambda(u \pm i\sqrt{v^2+w^2})}\,.
\]
The significance of the functions $u \pm i\sqrt{v^2+w^2}$ will be explained in the next section.

\bigskip

This first Hamiltonian structure may also, trivially, be obtained by taking the dispersionless
limit of the first Hamiltonian structure of the dispersive system (\ref{jordankdv}). It turns
out that this procedure fails when applied to the second Hamiltonian structure. Svinolupov
\cite{[Sv]} (see also \cite{[GK]}) found the recursion operator for (\ref{jordankdv}):
\[
{\cal R}^i_j = \delta^i_j \Bigg( \frac{d~}{dx}\Bigg)^2 +
\Bigg\{
    \frac{2}{3} a_{jk}^{~~i} u^k + \frac{1}{3} a_{jk}^{~~i} u_x^k
    \Bigg(\frac{d~}{dx}\Bigg)^{-1}
\Bigg\} +
\frac{1}{9}  \Delta_{kl}^{~~ji}
u^l \Bigg( \frac{d~}{dx} \Bigg)^{-1} \Bigg\{ u^k \Bigg( \frac{d~}{dx}\Bigg)^{-1}\Bigg\}\,.
\]
Applying this to the first Hamiltonian operator\footnote{The notation ${\cal A}_{(i)}$
will be used for dispersive Hamiltonian operators, and ${\cal H}_{(i)}$ for the dispersionless
limits of these operators.}
${\cal A}_{(1)}={\cal H}_{(1)}$
gives the second, compatible, Hamiltonian operator
\[
{\cal A}_{(2)}^{ij} = \eta^{ij} \Bigg( \frac{d~}{dx}\Bigg)^3 +
\Bigg\{
\frac{2}{3} a_{jk}^{~~i} u^k \Bigg(\frac{d~}{dx}\Bigg)+ \frac{1}{3} a_{jk}^{~~i} u_x^k
\Bigg\} +
\frac{1}{9} \Delta_{kl}^{~~ji} u^l  \Bigg( \frac{d~}{dx} \Bigg)^{-1} u^k\,.
\]
Note that
\[
\frac{\partial {\cal A}_{(2)}^{ij}}{\partial u^1} = \frac{2}{3} {\cal A}_{(1)}^{ij}\,.
\]
For ${\cal A}_{(2)}^{ij}$ to be a Hamiltonian operator requires that the structure
constants $c_{ij}^{~~k}$ define a Jordan algebra: up to now this property has not been used.

To perform the scaling
\begin{gather*}
\frac{d~}{dx}  \rightarrow  \epsilon \frac{d~}{dx} \,, \\
\frac{\partial~}{\partial t}  \rightarrow  \epsilon \frac{\partial~}{\partial t}
\end{gather*}
(under which the Euler operator
\[
\frac{\delta~}{\delta u^i} = \sum_{n=0}^\infty (-1)^n \Bigg( \frac{d~}{dx} \Bigg)^n
\frac{\partial~}{\partial u_{nx}}
\]
is invariant) one must expand an arbitrary Hamiltonian density as a power series in $\epsilon\,$
\[
H^{(n)} = h^{(n)} [{\bf u}] + \epsilon^2 \delta h^{(n)} [{\bf u},{\bf u_x},{\bf u_{xx}}]+
O(\epsilon^4)\,.
\]
so $h^{(n)}$ is purely hydrodynamic, $\delta h^{(n)} = \chi^{(n)}_{ij}({\bf u}) u^i_x u^j_x\,,$
etc.. Thus the dispersive system
\[
u^i_{t_n} = {\cal A}_{(2)}^{ij} \frac{\delta H^{(n)} }{\delta u^j}
\]
transforms to
\begin{equation}
u^i_{t_n} =
\Bigg\{
\frac{2}{3} a^{ij}_k u^k \frac{d~}{dx} + \frac{1}{3} a^{ij}_k u_x^k
\Bigg\}  \frac{\delta h^{(n)}}{\delta u^j} + \frac{1}{9}
\Delta_{mn}^{~~ji} u^n \Bigg( \frac{d~}{dx} \Bigg)^{-1}
\Bigg\{ u^m \frac{\delta \, \delta h^{(n)}}{\delta u^j}\Bigg\}\,.
\label{secondham}
\end{equation}
There is a potential singular term,
\[
\frac{1}{9} \lim_{\epsilon \rightarrow 0} \frac{1}{\epsilon^2} u^n
\frac{d~}{dx} \Bigg\{ \Delta_{mn}^{~~ji} u^m \frac{ \delta h^{(n)} }{\delta u^j}\Bigg\}
\]
but it may be shown that this vanishes when $h^{(n)}$ is a conserved density.
The form of this suggests one should define a metric and a connection by the
formulae
\begin{gather*}
g^{ij}  =  \frac{2}{3} c^{ij}_{~~k} u^k\,,\\
\Gamma^{ij}_k  =  \frac{1}{3} c^{ij}_{~~k}\quad\quad\quad
({\rm and~}\Gamma^{ij}_k = - g^{ir} \Gamma^i_{kr})\,.
\end{gather*}
In terms of the Jordan manifold structure the metric may be written invariantly as
\[
g^{ij} = \frac{2}{3} E( du^i \circ du^j)\,.
\]
Here the metric $\eta^{ij}$ has been used to induce a
multiplication, also denoted by $\circ\,,$ dual to
(\ref{tangentmult}) on each cotangent space of the manifold.
This metric has the property, derived from the existence of a unity in the Jordan algebra,
that
\[
{\cal L}_e g^{ij} = \frac{2}{3} \eta^{ij}\,.
\]
These formulae define a metric connection, since
\begin{align*}
\nabla_i g^{jk} & =  \partial_i g^{jk} + \Gamma_{ip}^j g^{pk} + \Gamma_{ip}^k g^{jp}\,,\\
& =  \partial_i g^{jk} - \Gamma_i^{jk} - \Gamma_i^{kj} \,, \\
& =  0\,.
\end{align*}
However the metric has torsion, the torsion tensor being the anti-symmetric part of the
connection $T^{i}_{jk} = \Gamma^i_{jk} - \Gamma^i_{kj}\,.$
This is related to the failure of the algebra defined by $c_{ij}^{~~k}$ to be
associative\footnote{Indices and $c_{ijk}$ and $\Delta_{ijk}^{~~~k}$
will always raises and lowered using $\eta^{ij}$ and its inverse, NOT by $g^{ij}$ and its
inverse.}:
\begin{align*}
T^{ijk} & =  g^{jp} g^{kq} T^i_{pq} \,,\\
& =  g^{kq} \Gamma^{ij}_q - g^{jq} \Gamma^{ik}_q\,,\\
& =  \frac{2}{9} \Delta^{jik}_{~~~~p} u^p\,.
\end{align*}
Similarly, the curvature also depends on the associator. Thus in the
case where the algebra is associative, where one has a Frobenius manifold, these structures
reduces to a flat, torsion free metric connection, as required by the Dubrovin-Novikov theorem \cite{[DN]}.

Because of the presence of torsion $(\nabla_i\nabla_j - \nabla_j\nabla_i) \phi \neq 0\,$ in
general. However, using the recursion relation (\ref{recursion}) one may show that
$(\nabla_i\nabla_j - \nabla_j\nabla_i) h^{(n)} =0\,.$ This result is behind the vanishing
of the potentially singular terms in (\ref{secondham}).

The presence of the second, non-local term in (\ref{secondham}) is an unusual feature. It
is non-local, but acts on $\delta h^{(n)}$ which is quadratic in derivatives. These two effects
conspire to produce a purely hydrodynamic flow. It may be shown, using ideas developed
for studying genus-one deformations in topological field theory \cite{[EYY]}, that the $\chi^{(n)}_{ij}$
satisfy the recursion relation \cite{[St]}
\begin{equation}
n\chi^{(n)}_{ij} = a_{ir}^{~~s} u^r \chi^{(n-1)}_{sj} - \frac{3}{2}
\frac{\partial^2 h^{(n-1)}}{\partial u^i \partial u^j}\,,
\label{firstorderrecursion}
\end{equation}
with initial condition $\chi^{(2)}_{ij} = 0\,.$ Thus $\chi^{(3)}_{ij}$ is a function of
$h^{(2)}$ alone, and hence, by recursion, $\chi^{(n)}_{ij}$ is a complicated function
of the densities $h^{(n)}\,$ and their derivatives.

\section{Riemann invariants}

A necessary and sufficient condition for a hydrodynamic system (\ref{genhydro}) to be put
into Riemann invariant form
\[
R^i_t = v^i[{\bf R}] R^i_x
\]
is the vanishing of the Haantjes tensor \cite{[N],[H]}. This is defined in terms of the
Nijenhuis tensor
\[
N^i_{jk} = v^p_j \partial_p v^i_k - v^p_k \partial_p v^i_j -
v^i_p(\partial_j v^p_k - \partial_k v^p_j)
\]
by
\[
T^i_{jk}= N^i_{pr} v_j^p v^r_k - N^p_{jr} v^i_p v^r_k - N^p_{rk} v^i_p v^r_j + N^p_{jk} v^i_r v^r_p\,.
\]
If this vanishes then the hydrodynamic system is integrable by the generalized hodograph
transform \cite{[T]}.

For the dispersionless KdV system the Nijenhuis tensor is linear;
\[
N^i_{jk} = \Delta_{jrk}^{~~~i} u^r\,,
\]
so for any system related to a Frobenius manifold, for which the associator vanishes, the
Haantjes tensor vanishes trivially. This reproduces a special case of the result that the
hydrodynamic systems obtained from Frobenius manifolds are integrable by the generalized
hodograph transformation. The vanishing of the Haantjes tensor is a quartic condition on the
structure constants. Jordan algebras may, or may not, have vanishing Haantjes tensor.

\bigskip

\noindent{\bf Example}. For the Jordan algebra constructed in Section 2 for $3\times 3$ matrices,
the corresponding
Haantjes tensor is non-zero. This may be shown by direct computation.

\bigskip

\noindent{\bf Example}. For the $D_N$-Jordan algebras the corresponding Haantjes tensor
vanishes. Using the representation of the structure constants (\ref{dnconstants}) one may show
\[
N^i_{jk} = \left\{ \begin{array}{ll}
0&{\rm if~any~}i\,,j\,,k=1 \\
u^k\delta_{ij} - u^j \delta_{ik} &{\rm otherwise} \end{array}\right.
\]
(In general, if the Jordan algebra has a unity then $\Delta_{ijk}^{~~~s} = 0$ if any
$i\,,j\,,k=1\,.$) Tedious but straightforward calculations then show that the
Haantjes tensor vanishes. Thus the $D_N$-Jordan algebra dispersionless KdV equations
may be written in Riemann invariant form.

\bigskip

\noindent{\bf Example}. ($D_3$-Jordan algebra)

In terms of the Riemann invariants
\begin{gather*}
R^1  =  u + \sqrt{v^2+w^2}\,,\\
R^2  =  u - \sqrt{v^2+w^2}\,,\\
R^3  =  (v^2+w^2)/(vw)
\end{gather*}
the system (\ref{basicexample}) becomes (to ensure a hyperbolic system the transformations
$v\rightarrow i v\,, w\rightarrow i w$ have been made):
\begin{equation}
\begin{split}
R^1_t & =  R^1 R^1_x\,, \\
R^2_t & =  R^2 R^2_x\,, \\
R^3_t & =  (R^1+R^2)/2 \,\,R^3_x\,.
\end{split}
\label{rd3}
\end{equation}
This shows a certain degree of degeneracy: the first two equation are completely
decoupled, and the third is linear. This does not contradict Svinolupov's result
on the irreducibility of the corresponding KdV system, only linear transformation
are allowed for KdV systems -- nonlinear ones destroying the form of the equations,
while for dispersionless systems nonlinear transformation preserve the form
of a hydrodynamic system. Furthermore,
the conservation laws constructed earlier are independent of $R^3\,,$ in fact they too
decouple:
\[
h^{(n)} = \frac{1}{2n!} ({R^1}^n + {R^2}^n)\,.
\]
More generally any function $q_1(R^1)+q_2(R^2)$ is a conserved density for this system.

\section{Conclusions}

The Jordan KdV equations have all the properties one would expect from a completely
integrable system -- even though Lax equations for them have not been found in all cases.
This paper is a first attempt to study the dispersionless limits of these systems.
The obvious question is whether (and in what sense) the systems remain integrable in this limit.
For any $3$-component Hamiltonian system of hydrodynamic type one has the following result:

\medskip

\noindent{\bf Theorem}. \cite{[F1]}

\medskip

\noindent A $3$-component Hamiltonian system is integrable if and
only if one of the following conditions are fulfilled:

\begin{itemize}

\item[$\bullet$] the system is diagonalizable (and hence integrable by the generalized
hodograph transform);

\item[$\bullet$] the system is non-diagonalizable, but weakly
non-linear (and hence integrable by the inverse scattering
transform).

\end{itemize}

\noindent For systems with more components there are no analogues to
this theorem, though certain conjectures have been made \cite{[F2]}.
For certain systems, such as the $D_3$-Jordan system, the system is semi-Hamiltonian and hence
integrable by the generalized hodograph transform. However not all systems are semi-Hamiltonian,
or even have maximal numbers of Riemann invariants, and for such systems the precise
nature of their integrability remains unclear.

Properties of conserved densities and the first Hamiltonian structure of these systems
remain unchanged under this limit. The existence of the unity element in the Jordan algebra,
one of the additional assumptions made in this paper, enables
conserved densities to be defined recursively. However, direct averaging of
the second Hamiltonian structure is problematic. One obtains a non-local operator which depends
both on $h^{(n)}$ and $\delta h^{(n)}\,,$ so there remains a vestige of the original dispersive
system. The term $\delta h^{(n)}$ may, via (\ref{firstorderrecursion}), be calculated
in terms of $h^{(n)}\,,$ so in an implicit way the second structure ${\cal H}_{(2)}$ is a function
of the purely hydrodynamic part of the dispersive Hamiltonian density. All these problems
disappear if the algebra is associative, where one reproduces standard results from the
theory of Frobenius manifolds.

There are many interesting systems which fall outside, though are closely related to,
the class considered here. Probably the
simplest is the dispersionless limit of Ito's system \cite{[I]}, which
falls into a class of equations of KdV-type associated to any Lie
algebra \cite{[K]}:
\begin{gather*}
u_t  =  3 u u_x + v v_x \,, \\
v_t  =  (u v)_x\,.
\end{gather*}
This is of the general form (\ref{djordankdv}), though the algebra defined by the
structure constants is not a Jordan algebra. This system is
bi-Hamiltonian with operators
\begin{gather*}
{\cal H}_{(1)}  =  \left(\begin{array}{cc} D & 0 \\ 0 & D \end{array}\right) \,, \\
{\cal H}_{(2)}  =  \left(\begin{array}{cc} u D + D u & vD \\ D v & 0
\end{array}\right)\,,
\end{gather*}
and in this case these are the dispersionless limits of the corresponding dispersive
operators \cite{[Do]}. The dispersive
counterpart is of the general form
\[
u^i_t = b^{i}_j u^j_{xxx} + a_{jk}^{~~i} u^j u^k_x\,,
\]
where $b^i_j$ is a degenerate, constant, matrix. Clearly further work is required to
understand the properties of
compatible Hamiltonian structures under the dispersionless limit,
both for the Jordan systems \cite{[Sv]} and for the Lie-algebra systems \cite{[K]}.

An alternative approach is to use the hydrodynamic non-local
Hamiltonian operators, as developed by Ferapontov (see appendix).
This has the advantage of being comparatively simple to apply to a
specific system, though in moving to Riemann invariant form (if
such a form exists) the connection with Jordan algebras is lost
and so it seems hard to develop a general theory of dispersionless
KdV equations along these lines. The example of the $D_3$-Jordan
system shows that the resulting structures may be somewhat
degenerate; two of the equations decouple and the third is linear.
For $N\geq 4$ the $D_N$-Jordan systems have repeated eigenvalues
(or characteristic velocities). The appearance of repeated
characteristic velocities, where $v^i[{\bf R}]=v^j[{\bf R}]$ for
certain $i\neq j$ makes finding the metric more problematical,
though not impossible. Other examples with such repeated
characteristic velocities, also related to Jordan algebras, have
been studied in \cite{[Mc],[Mcthesis]}.

\subsection*{Acknowledgments}
I would like to thank Oscar McCarthy for numerous discussions on the topic of
Jordan KdV systems.

\section*{Appendix: Nonlocal Hamiltonian Operators}

In this appendix we restrict our attention to the system (\ref{rd3}). The Hamiltonian
structure for this will involve non-local tails, that is, be of the form
\[
A^{ij} = g^{ij}\frac{d~}{dx} - g^{is} \Gamma^j_{sk} u^k_x +
\sum_\alpha  \omega_{\alpha k}^i u^k_x \left(\frac{d~}{dx}\right)^{-1}
\omega_{\alpha l} u^l_x\,,
\]
the conditions for this to be Hamiltonian being given in \cite{[F3]}. The system (\ref{rd3}) is
semi-Hamiltonian, the characteristic speeds satisfying the system
\[
\partial_k \Bigg\{ \frac{ \partial_j v^i}{v^j-v_i} \Bigg\} =
\partial_j \Bigg\{ \frac{ \partial_k v^i}{v^k-v_i} \Bigg\} \quad\quad
\forall i\neq j\neq k \neq i\,.
\]
The Hamiltonian structure is given in terms of a diagonal metric
${\rm g} = \sum g_{ii} d{R^i}^2\,,$ where the components are defined by
the equation
\[
\partial_j \log \sqrt{g_{ii}} =  \frac{ \partial_j v^i}{v^j-v_i} \quad\quad\forall i\neq j\,,
\]
the semi-Hamiltonian condition being the integrability condition for this system. For
the $D_3$-dispersionless Jordan system this metric is
\[
{\bf g} = \frac{d{R^1}^2}{\phi_1(R^1)} + \frac{d{R^2}^2}{\phi_2(R^2)}+
\frac{(R^1-R^2)^2}{\phi_3(R^3)}d{R^3}^2\,,
\]
where $\phi_i$ are arbitrary functions of their arguments. Commuting flows are given
by $R^i_\tau = \omega^i[{\bf R}] R^i_x\,,$ where the functions $w^i$ solve the system
\[
\frac{ \partial_j w^i}{w^j-w_i}=\frac{ \partial_j v^i}{v^j-v_i} \quad\quad\forall i\neq j\,.
\]
For this system the solutions are easily obtained:
\begin{gather*}
R^1_\tau  =  \psi_1^\prime(R^1) R^1_x \,, \\
R^2_\tau  =  \psi_2^\prime(R^2) R^2_x \,, \\
R^3_\tau  =  \Bigg[
\frac{ \psi_1(R^1) - \psi_2(R^2) + \psi_3(R^3) }{R^1-R^2} \Bigg] R^3_x\,.
\end{gather*}
Note that for this system the commuting flows are labeled by three arbitrary functions
$\psi_i$ rather than a discrete label. This is already manifest in the dispersionless
KdV equation, where the flows
\begin{gather*}
u_t  =  u u_x \,, \\
u_\tau  =  f(u) u_x
\end{gather*}
commute for all functions $f(u)$, not just for the functions $f(u)=u^n$ which
correspond to the dispersionless limits of the full KdV hierarchy.

The functional dependence in the Weingarten operators $\omega^i_\alpha$ is now fixed
by requiring that the components $R^{ij}_{ij}=g^{ii} R^j_{iij}$ satisfy the
equation
\[
R^{ij}_{ij}=\sum_\alpha \varepsilon_\alpha \omega^i_\alpha w^j_\alpha\,, \quad\quad
\varepsilon=\pm 1\,.
\]
Here $\alpha$ labels the functional dependence in Weingarten operators. This expansion
fixes these in terms of the arbitrary functions in the metric.

Thus to each set of functions $\phi^i$ one obtains a non-local
Hamiltonian operator. These are all mutually compatible, resulting
in a multi-Hamiltonian structures. The full details of the
multi-Hamiltonian structure of the $D_N$-dispersionless Jordan
systems may be found in \cite{[Mcthesis]}.

\label{lastpage}

\end{document}